\documentclass[11pt,twoside]{article}

\usepackage{asp2006}
\usepackage{epsf}
\usepackage{psfig}
\usepackage{lscape}

\begin{document}
\title{Evolution of the Red Sequence in simulated galaxy groups}   
\author{Alessio D. Romeo}   
\author{Nicola R. Napolitano, Giovanni Covone}
\affil{INAF - Osservatorio Astronomico di Capodimonte, v. Moiariello 16, I-80131 Napoli, Italy}    
\author{Jesper Sommer-Larsen}
\affil{Dark Cosmology Centre, Niels Bohr Institute, University of Copenhagen, J. Maries v. 30, DK-2100
Copenhagen, Denmark}

\begin{abstract} 
N--body + hydrodynamical simulations of formation and evolution
of galaxy groups in a $\Lambda$CDM cosmology have been performed.
The properties of the galaxy populations in 12
groups are here discussed, with focus on the colour-magnitude relation
in both normal and fossil groups.
\end{abstract}


\section{Introduction}
The light and stellar mass in clusters of galaxies is dominated by bright,
massive ellipticals, which form a tight colour-magnitude relation
\citep[CMr:][]{bower92,gladders}.
The CMr is classically interpreted as a 
mass--metallicity relation \citep{KA}, its slope being mainly driven by
the latter: brighter galaxies are redder because more metal enriched.
As for groups of galaxies, they are by far less studied both observationally and theoretically 
than clusters.
Our aim is to study, by means of cosmological and hydrodynamical simulations,
the CMr in a wider range of both mass and redshift, including groups, cluster cores and cluster
outskirts from $z$=2 to $z$=0
\citep[for simulations or semi-analytical models of cluster galaxies see][]{kaviraj,saro}.
This approach allows to investigate how the relation evolves with time and can
shed light on possible sistematic differences due to intermediate
density environment \citep[see also][]{tanaka}.

The groups were drawn from a Tree-SPH simulation, within a standard flat $\Lambda$CDM 
cosmological model, which consistently includes star formation, chemical evolution with
non-instantaneous recycling, metal dependent radiative cooling, strong 
starbursts and galactic super winds:
full details on the code and the simulations are given in \citet{PI}.
In particular we selected 12 groups (all of $T\sim 1.5$ keV), four of which turned out to be 
fossil \citep[FGs: see][]{DO}.

\begin{figure}[!h]
\plotone{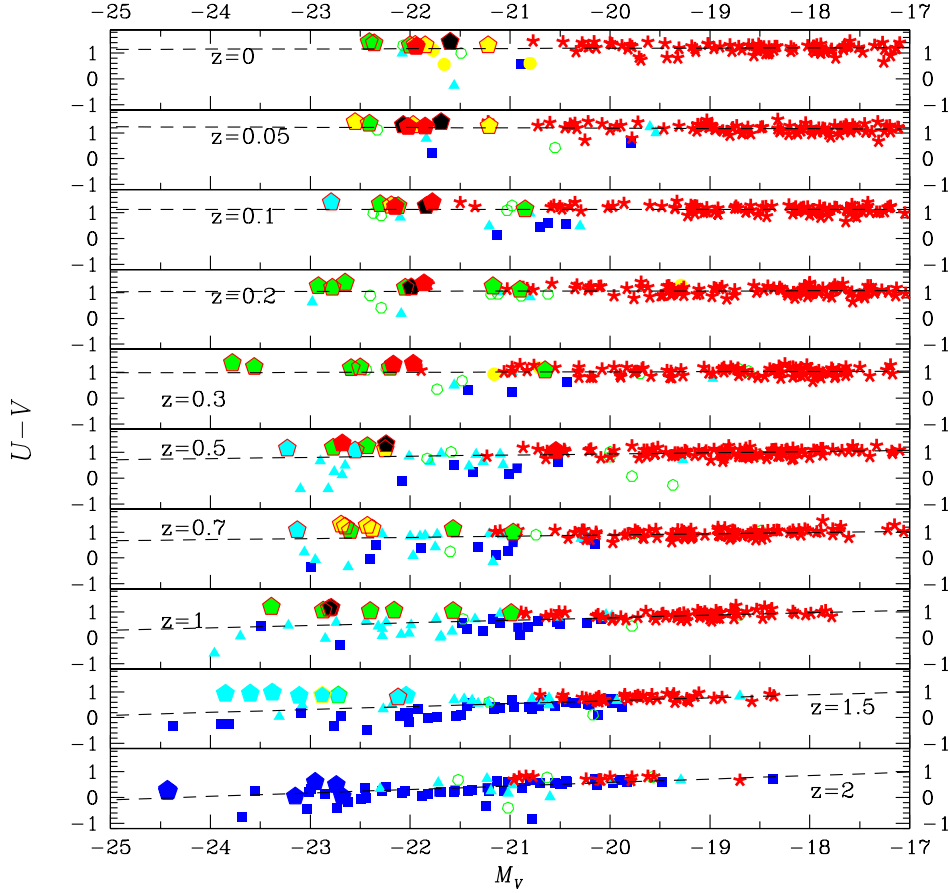}
  \caption{
  {\it U-V} vs. {\it V} CMr for normal groups, at different
  redshifts. Galaxies are coloured by their specific star-formation rate (SSFR): 
  $<$0.01 (red stars), 0.01--0.1 (yellow filled circles), 0.1--1 (green open circles), 
  1--5 (cyan triangles), $>$5 (blue squares) in units of
 $Gyr^{-1}$/10. Central brightest galaxies are marked as pentagons (see text).
  }
\label{}                  
\end{figure}

\begin{figure}[!h]
\plotone{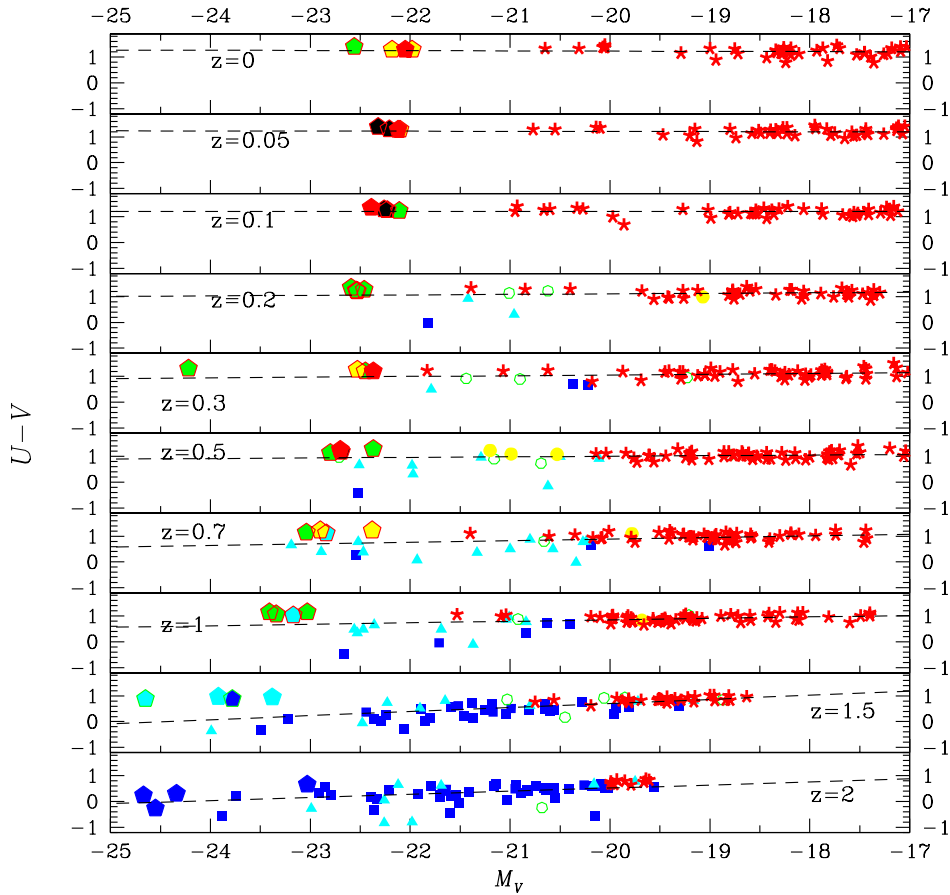}
  \caption{Same as previous Fig. for galaxies in fossil groups.}
\label{}                  
\end{figure}

\section{Colour-magnitude relation of simulated groups}

The main quantitative information about the CMr are given by its slope, zero-point
and slope. 
The zero-point constrains the epoch of the bulk of star formation, which is at around $z\sim$3 for most of the 
cluster ellipticals \citep{kodama98}. The scatter of the CMr is observed to be small in the local Universe 
\citep[$\sim$0.04 in clusters, $\sim$0.1 in fields:][]{bower92,terlevich} and does not increase up to $z\sim$1 \citep{blakeslee}. 
Such persisting 
tightness of the CMr suggests that the bulk of stellar population have formed across a relatively short period, 
after which galaxies have evolved mostly driven by passive reddening \citep{ellis06,gladders}, in concordance with 
``down-sizing'' models of galaxy formation \citep[see][]{kodama04,cimatti06}. 
In a hierarchical scenario, on the contrary, the slope of the CMr is 
predicted to flatten with $z$, as a result of a merger-driven evolution of the largest ellipticals. 
\citet{delucia04} found a deficit of red, faint galaxies populating the CMr at $z$=0.8, indicating that
star-forming galaxies 
shifted towards the Red Sequence when their activity came to an end at around that redshift.

We are able to follow the evolution up to farther redshifts, to determine at which epoch, possibly, the 
scatter becomes larger and the slope changes, which is precisely the point when the Red Sequence is built up. 
In Figs. 1 and 2 the colour ({\it U-V})--magnitude ({\it V}) relation of galaxies in
the groups, respectively normal and fossil, is shown. Galaxies are classified according to
their specific star-formation rate (SSFR) over the last previous Gyr. The linear best-fits are calculated on a sub-sample
of galaxies, namely those laying within a region of the colour -- internal velocity dispersion space,
which can be consistent with early-type objects.

By comparing Figs. 1 and 2, one can follow the assembly process of the brightest central galaxies (BCG)
with respect to the `fossilness' of 
the group: star formation activity lasts longer and the number of star forming galaxies remains higher in normal groups than in 
fossil ones, supporting the picture of the latters being rather ``quiescent", earlier assembled systems: in FGs the gap between the 
BCG and the 2nd brightest galaxy opens already since $z\sim$1.5 and then widens out.
 
In Fig. 3 the slope and the scatter of the CMr from the galaxies more massive than $2\cdot 10^{10}M_{\odot}$ 
are shown, along with the same for the other two classes
of environments considered, that is the core (within 1/3 $R_{vir}$) and the outskirts (from 1/3 $R_{vir}$ to $R_{vir}$)
of two clusters. 
The tightness of the CMr spreads up from $z\sim$1 onwards for clusters, $\sim$0.5 for groups: at lower redshifts the scatter keeps 
a slightly increasing trend, for all 
the classes considered and especially in groups, staying within 0.2 mag.  
This is mirrored by the slope's behaviour: in groups the CMr remains almost
flat at low $z$, becomes positive at later $z$ with respect to cluster cores (where $z\sim$0.7) and keeps on steeper at higher $z$,
following an evolution rather analogous to cluster outskirt regions than cluster cores. 

In conclusion, whereas the main difference in buiding the Red Sequence between normal and fossil groups stems from
the different timescales of star formation activity and in the mass assembly of the central galaxy, with respect to clusters 
the CMr seems to evolve with a slowed down pace: galaxies in groups are still in the phase of moving towards
the Red Sequence at epochs when the latter is already in place for galaxies in clusters; the most
marked difference is found between normal groups and cluster cores.


\begin{figure}
\plottwo{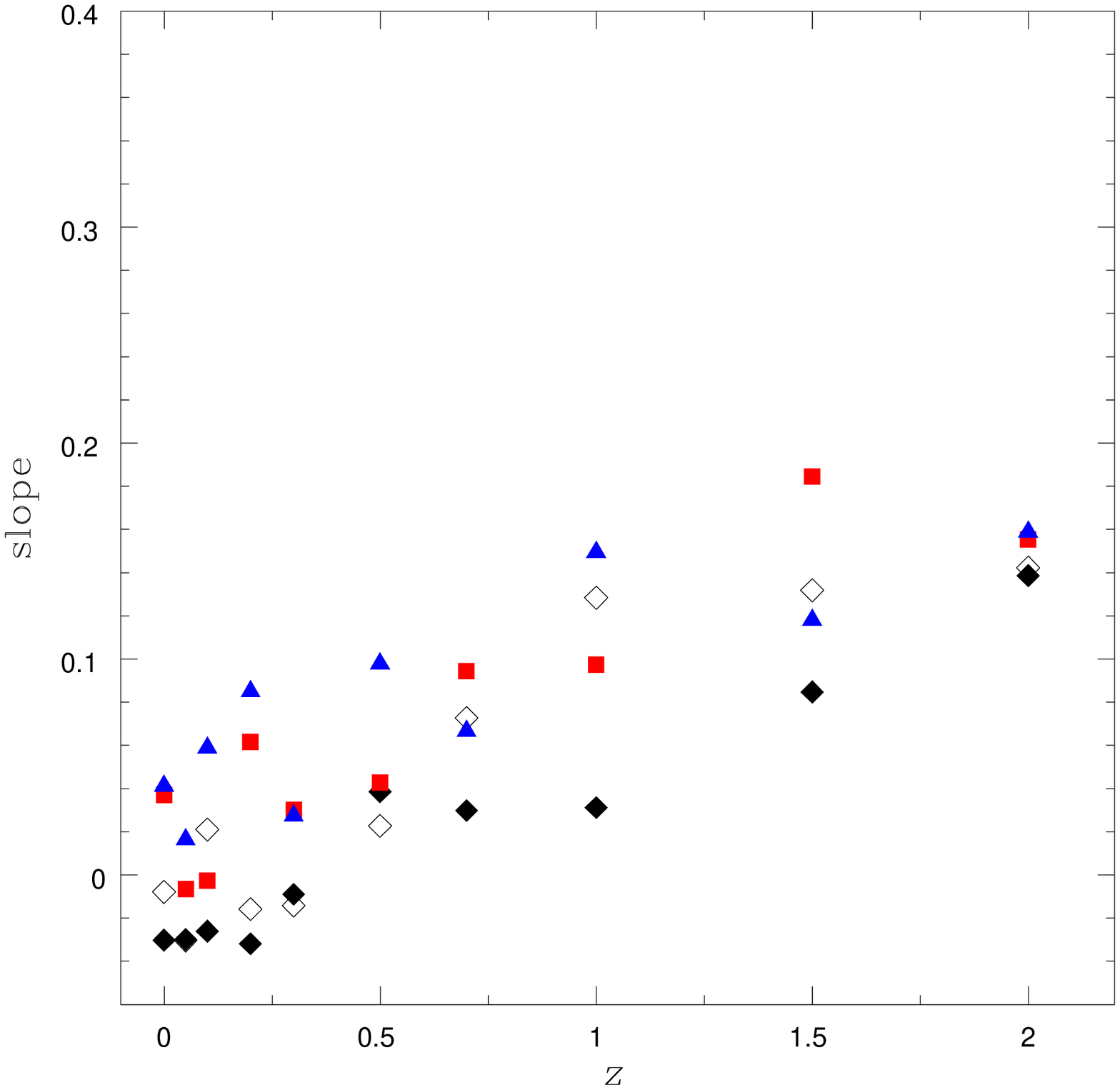}{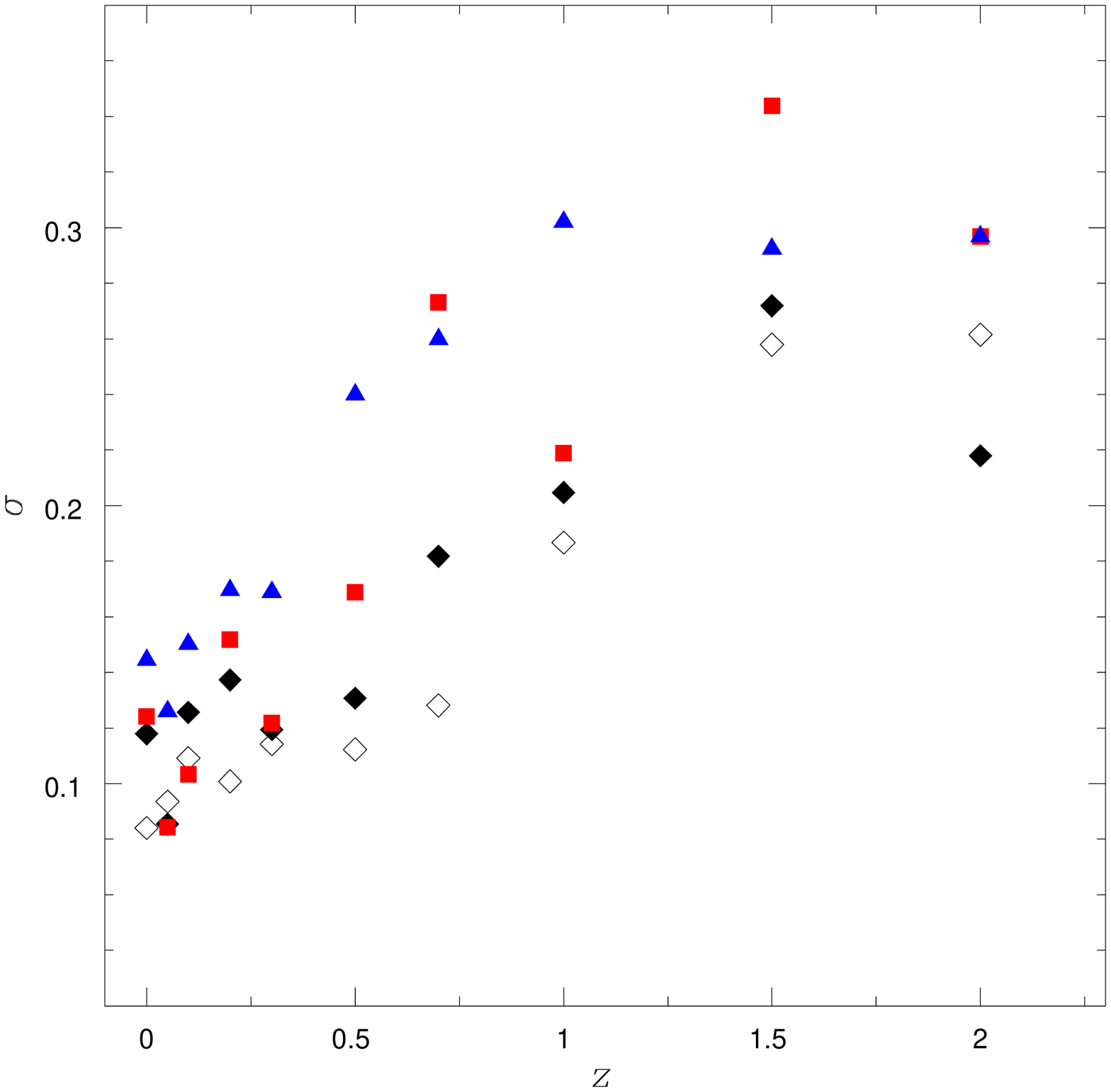}
\caption{Slope of the linear fit ({\it left}) and
2-$\sigma$ scatter ({\it right}) of the CMr for
   galaxies more massive than $2\cdot 10^{10}M_{\odot}$ in the cluster cores 
  (black diamonds), cluster outskirts (open diamonds), normal groups
  (blue triangles) and fossil groups (red squares).}
\label{}                  
\end{figure}




\end{document}